\documentclass[final,3p,times,twocolumn]{elsarticle}

\usepackage{graphicx}
\usepackage{slashed}
\usepackage{amssymb}

\usepackage{lineno}
\usepackage{verbatim} 

\makeatletter
\def\ps@pprintTitle{%
 \let\@oddhead\@empty
 \let\@evenhead\@empty
 \def\@oddfoot{}%
 \let\@evenfoot\@oddfoot}
\makeatother
\begin{document}

\begin{frontmatter}


\title{Constraining Stealth SUSY with illuminated fat jets at the LHC}



\author[label1,label3]{Marvin Flores}
\ead{mflores@nip.upd.edu.ph}
\author[label1]{Deepak Kar}
\author[label2]{Jong Soo Kim}

\address[label1]{School of Physics, University of the Witwatersrand, Johannesburg, South Africa}
\address[label2]{National Institute for Theoretical Physics and School of Physics, University of the Witwatersrand, Johannesburg, South Africa}
\address[label3]{National Institute of Physics, University of the Philippines, Diliman, Quezon City, Philippines}

\begin{abstract}
We investigate the discovery potential of a Stealth SUSY scenario involving squark decays by reconstructing the lightest neutralino decay products using a large-radius jet containing a high transverse momentum photon. Requirements on the event topology, such as photon and large-radius jet multiplicity result in less background than signal. We also estimated the sensitivity of our analysis and found that it has a better exclusion potential compared to the strongest existing search for the specific benchmark points considered here.
\end{abstract}


\end{frontmatter}


\section{Introduction}
\label{S:1}

Among the existing beyond-the-Standard-Model (BSM) scenarios, supersymmetry (SUSY) is the leading theoretical framework that explains unresolved questions in the Standard Model such as the large hierarchy between the weak scale and the Planck scale \cite{susy,dreessusy,nillessusy,gunionsusy}. Hence, it is extensively being probed at the Large Hadron Collider (LHC). However, after collecting data for more than eight years, no searches in favor of SUSY has yet been found \cite{experiment1,experiment2,experiment3,experiment4,experiment5,experiment6}. Even in full models, the limits on the SUSY parameter space are rather strict \cite{Drees:2015aeo,Kim:2016rsd,Domingo:2018ykx, gambit1, gambit2, fittino1, fittino2, mastercode,Bertone:2015tza}. As a consequence, some proponents of SUSY begin to wonder if the framework should be abandoned altogether.

In order not to abandon the appealing ideas of SUSY, models known as \textit{Stealth Supersymmetry} were introduced so as to evade existing standard SUSY searches \cite{stealth}. These searches typically rely on a large amount of missing transverse energy ($\slashed{E}_T$) \cite{experiment1,experiment2,experiment3,experiment4}, an approach motivated by $R$-parity which when preserved means that the lightest superpartner (LSP) is stable and contributes to missing energy. Therefore, Stealth SUSY scenarios seeks to reduce this $\slashed{E}_T$ as much as possible \cite{stealth, stealth2, stealth3}. 

The simplest Stealth SUSY model does this by making the standard LSP take on a new role as the lightest ``visible sector'' SUSY particle (LVSP) which decays into a lighter \textit{hidden} sector SUSY particle. The mass configuration is setup such that the boson and fermion of the hidden chiral supermultiplet are almost degenerate so that when the former decays to the latter, it leaves little phase space for the true LSP to carry energy, thereby producing signatures of low $\slashed{E}_T$.

In this paper, we study the above scenario by considering a particular toy model with a specific decay chain given by

\begin{equation}
\label{eq:decay}
\tilde{q} \rightarrow q(\tilde{\chi}_1^0 \rightarrow \gamma(\tilde{S} \rightarrow \tilde{G}(S \rightarrow gg))),
\end{equation}

\noindent where $\tilde{\chi}_1^0$, the lightest neutralino (hereafter referred to as a ``bino''), plays the role of our LVSP, and $\tilde{S}$ is our hidden SUSY particle, being the fermionic ``singlino'' superpartner of the singlet $S$, with the gravitino $\tilde{G}$ playing the role of our LSP. The presence of a high-$p_T$ photon is significant since it allows us to reconstruct the bino LVSP peak by searching for a pair of large-radius jets containing a high-transverse momentum photon, that is, $m_{\tilde{\chi}_1^0} \approx M(\gamma gg)$, something that has not been used in experiments so far. This was first pointed out in Ref. \cite{stealth} whereby ``illuminating'' a jet (i.e., having a high-$p_T$ photon inside it) renders the stealthiness weaker. We improve upon it by looking at specific event topology and show that imposing additional requirements such as large-radius jet and photon multiplicity result in less background than signal and consequently even stronger exclusion potential, thereby encouraging low-$\slashed{E}_T$ searches as promising alternatives to the usual high-$\slashed{E}_T$ ones that were performed at the LHC to look for SUSY.

During the preparation of this paper, CMS released a preliminary result \cite{cms3} that targets the same final state we are considering (i.e., $\gamma gg$) in a large-radius jet similar to the analysis employed in this paper although they considered gluino pair production instead of squarks that we consider here. This marks the third LHC search dedicated in probing Stealth SUSY. Whereas the former two searches relied on isolated photons \cite{cms, cms2}, the most recent one now relies on collimated photons and gluons.

This paper is arranged as follows: Section \ref{S:2} introduces the theory behind Stealth SUSY models and also motivates our specific toy model. Section \ref{S:3} then explains the details of the numerical simulations and analysis as well as our results involving the reconstructed bino LVSP at various benchmark points. Finally, we draw our conclusions in Section \ref{S:4}.

\section{Stealth Supersymmetry}
\label{S:2}

Many searches for new physics are reliant on large missing transverse energy ($\slashed{E}_T$) \cite{experiment1,experiment2,experiment3,experiment4} so a promising approach to avoid strong exclusion limits from these searches is to reduce $\slashed{E}_T$ as much as possible.

One such approach is R-parity violation \cite{Dreiner:1997uz,Barbier:2004ez} where the LSP is unstable and its decay products may be subject to detection. However, less missing transverse momentum is produced on average \cite{rpv,Dercks:2017lfq}.

On the other hand, we have R-parity-preserving models such as the so-called compressed SUSY models with little net $\slashed{E}_T$ \cite{Dreiner:2012gx,Carena:2008mj,Drees:2012dd}, as well as models known as Stealth SUSY which is a genuine reduction of $\slashed{E}_T$ due to having light LSP that carries little energy.

In this section, we discuss simplified stealth models relevant to our phenomenological study. More in-depth discussions can be found in \cite{stealth, stealth2, stealth3}. Stealth SUSY models typically involve the introduction of a hidden/stealth sector although it was pointed out in \cite{nmssm,jongstealth} that such a stealth sector is not needed. One could setup the necessary mass configuration in next-to-minimal supersymmetric Standard Model (NMSSM) by making the bino NLSP decay invisibly into a singlino LSP plus a singlet.

The main and crucial point with these stealth scenarios is that the hidden SUSY is almost unbroken. Thus the singlino and its singlet partner are mass degenerate, with the latter almost filling the mass gap between the singlino and the LSP. As a consequence, very little $\slashed{E}_T$ will be expected.

In order to achieve the stealth mechanism, we follow the model in \cite{stealth} and imagine that the LVSP can decay to a hidden sector field via some portal. Then, a decay chain within the hidden sector can occur ending with a massive $R$-odd stealth particle decaying to a nearly degenerate $R$-even state plus a light $R$-odd state. The $R$-even state must then decay into visible SM particles. In the simplest case, the hidden sector is taken to be a gauge singlet multiplet with a fermion $\tilde{S}$ and an almost degenerate scalar $S$ while the lightest superparticle in the spectrum is a gravitino.

\begin{figure}[h]
\centering\includegraphics[width=1.0\linewidth]{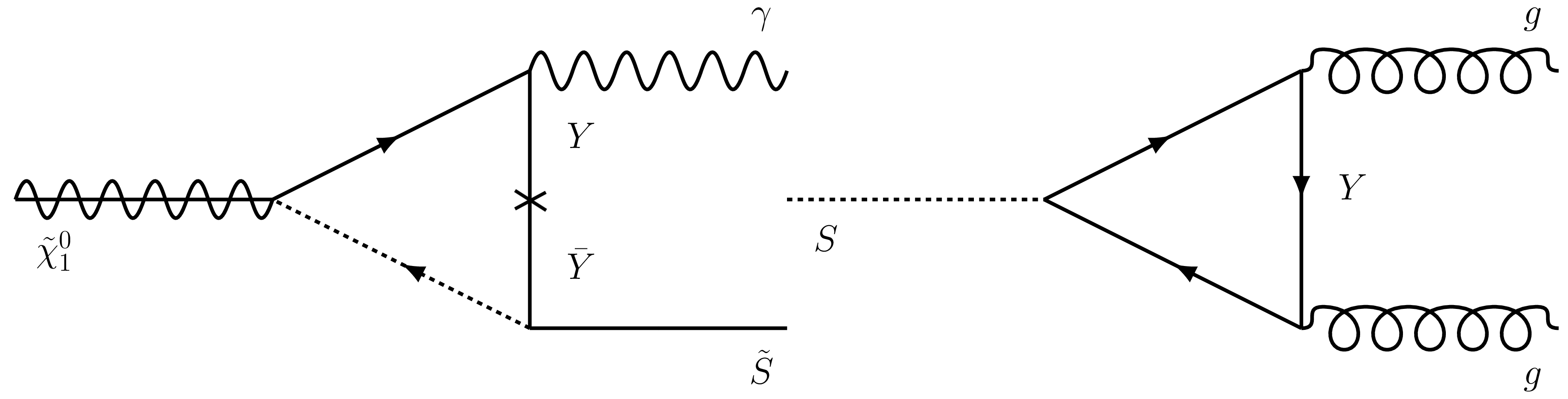}
\caption{Loop-induced couplings with vector-like states allowing the decay of a bino LVSP to a singlino and photon as well as the decay of singlet scalar to gluons.}
\label{fig:portals}
\end{figure}

One appropriate portal for the LVSP (in our case taken to be the bino) going to the stealth sector to proceed is via vector-like states $Y$, $\bar{Y}$ charged under SM and a $SY\bar{Y}$ coupling as discussed in \cite{stealth2} through the superpotential

\begin{equation}
W = \frac{m}{2}S^2 + \lambda S Y \bar{Y} + m_Y Y \bar{Y}
\end{equation}

\noindent where $m$ and $m_Y$ are the supersymmetric masses and $\lambda$ is the coupling between the singlet chiral superfield $S$ and the messenger field $Y$. This can induce a one-loop bino-photon-$\tilde{S}$ vertex allowing bino decays into $\tilde{S}$ while radiating off a photon, as well as inducing decays of a scalar $S$ to gluons as can be seen in Fig. \ref{fig:portals}.

\begin{figure}[h]
\centering\includegraphics[width=1.0\linewidth]{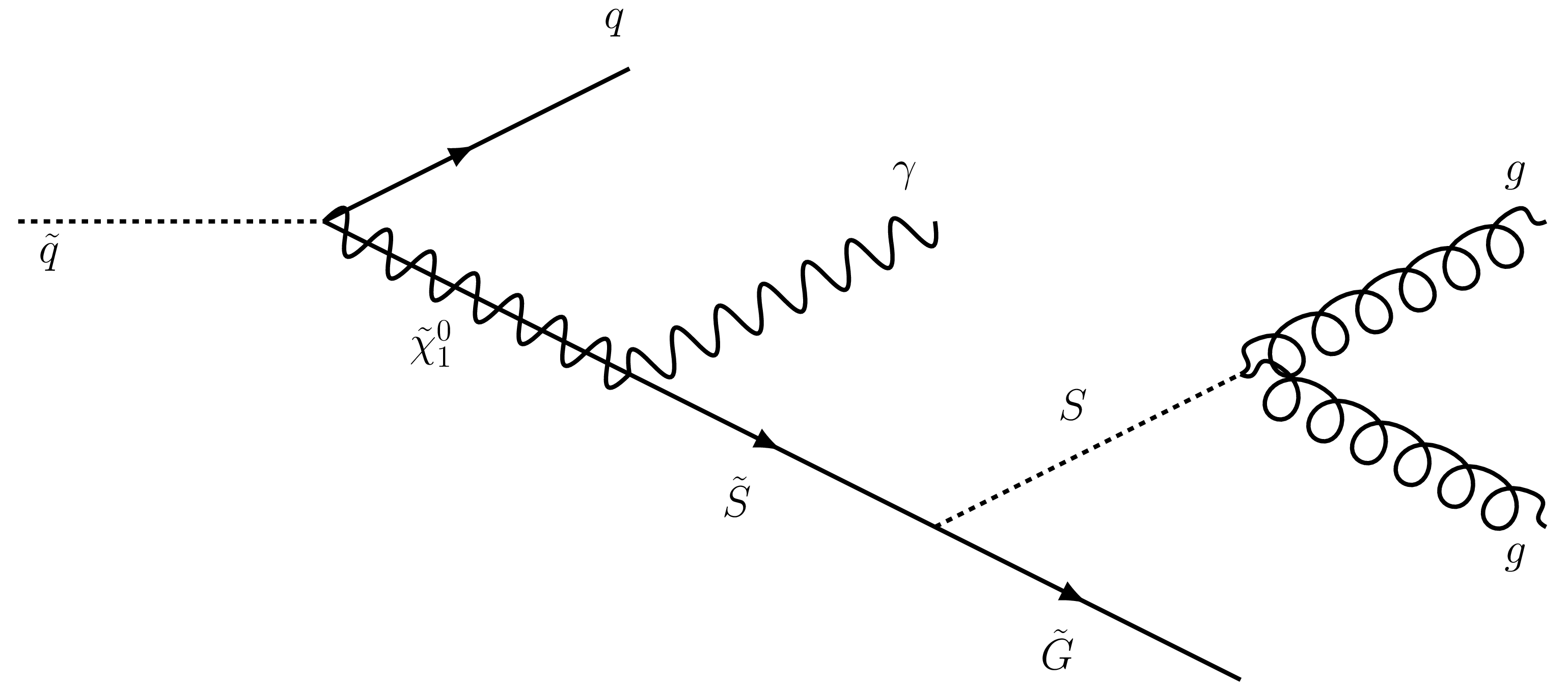}
\caption{The stealth decay chain considered in this study.}
\label{fig:decaychain}
\end{figure}

This grants us to consider a specific stealth decay chain shown in Fig. \ref{fig:decaychain} enabling us to search for resonances composed of a photon and a pair of jets arising from the gluons to reconstruct the bino LVSP.

It should be noted that initial attempts where made to reconstruct the squark itself via $M(\gamma g g q) \approx m_{\tilde{q}}$ but then we found that we would have to define jets with extremely large radius at around $R = 2.0$. Unfortunately, no experiments use large jets of this radius so we settled for the bino reconstruction instead.

\section{Numerical Analysis and Results}
\label{S:3}

We generated $20,000$ events for squark pair production with \textsc{Pythia 8.235} \cite{pythia} at the center-of-mass energy 13 TeV using the \textsc{NNPDF 2.3 QCD+QED LO} parton distribution function set \cite{pdf}. We used a modified decay table where we assumed a branching ratio\footnote{The branching ratios are intentionally simplified this way rather than calculated from a complete theory because we want to focus on the phenomenology of the final state $\gamma g g$ and see whether we can construct resonances from this specific topology. However, typical branching ratio of bino into photons and singlino can be of the order $10^{-3}$ \cite{stealth} but we assume this to be equal to unity for our simplified SUSY model since in certain region of parameter space BR($\tilde{\chi}_1^0 \rightarrow \gamma + \tilde{S}) \sim \mathcal{O}(1)$ is possible \cite{jongstealth}.} equal to $1$ for each of the decay in the chain given in Eq.~\ref{eq:decay} as well as having the following masses: the gluino mass is fixed at $m_{\tilde{g}} = 3000$ GeV while the squark mass $m_{\tilde{q}}$ is varied in steps of $50$ GeV from $1450$ to $2000$ GeV as well as the bino mass $m_{\tilde{\chi}_1^0}$ in steps of $50$ GeV from $250$ to $400$ GeV. The singlino and singlet masses are kept at $m_{\tilde{S}} = 100$ GeV and $m_S = 95$ GeV respectively with $\delta m = 5$ GeV. The respective production cross sections were obtained using \textsc{NNLLFast 1.1} \cite{nnllfast}, but reduced by a factor of $4/5$ since we are only considering the first two generations of squarks while \textsc{NNLLFast} sums over all flavours of final-state squarks including both chiralities, except for stops. The cross section of our signal ranges from $0.028$ pb for $m_{\tilde{q}} = 1450$ GeV down to $0.0033$ pb for $m_{\tilde{q}} = 2000$ GeV. \textsc{CheckMATE} \cite{checkmate,Kim:2015wza,Drees:2013wra} then tests all model points against existing LHC searches at $\sqrt{s} = 13$ TeV to see which benchmark scenarios evade them.

To determine whether a point is excluded by a search or not, \textsc{CheckMATE} compares the estimate of signal events with observed limits at $95\%$ C. L. of the search using

\begin{equation}
\label{eq:r}
r = \frac{s - 1.96\cdot\Delta s}{s^{95}_{\mbox{exp}}}
\end{equation}

\noindent where $s$ denotes the number of signal events, $\Delta s$ the uncertainty of MC events considered only to be the statistical uncertainty, $\Delta s = \sqrt{s}$. The value of $r$ is then calculated for every signal region of every search. In order to calculate the best exclusion limit, the ``best'' signal region is chosen as the one with the best expected exclusion potential. One can then define a point as excluded when $r > 1$. However, due to the fact that we do not control higher-order corrections or systematic errors, this calls for a definition of a region where exclusion is inconclusive and we define this to be the case when $0.67 < r < 1.5$. That is, when one of the points falls within the range of these $r$-values, we cannot tell whether it is excluded or allowed. Accordingly, we define a point as allowed whenever $r < 0.67$ and excluded when $r > 1.5$.

\begin{figure}[t]
\centering\includegraphics[width=1.0\linewidth]{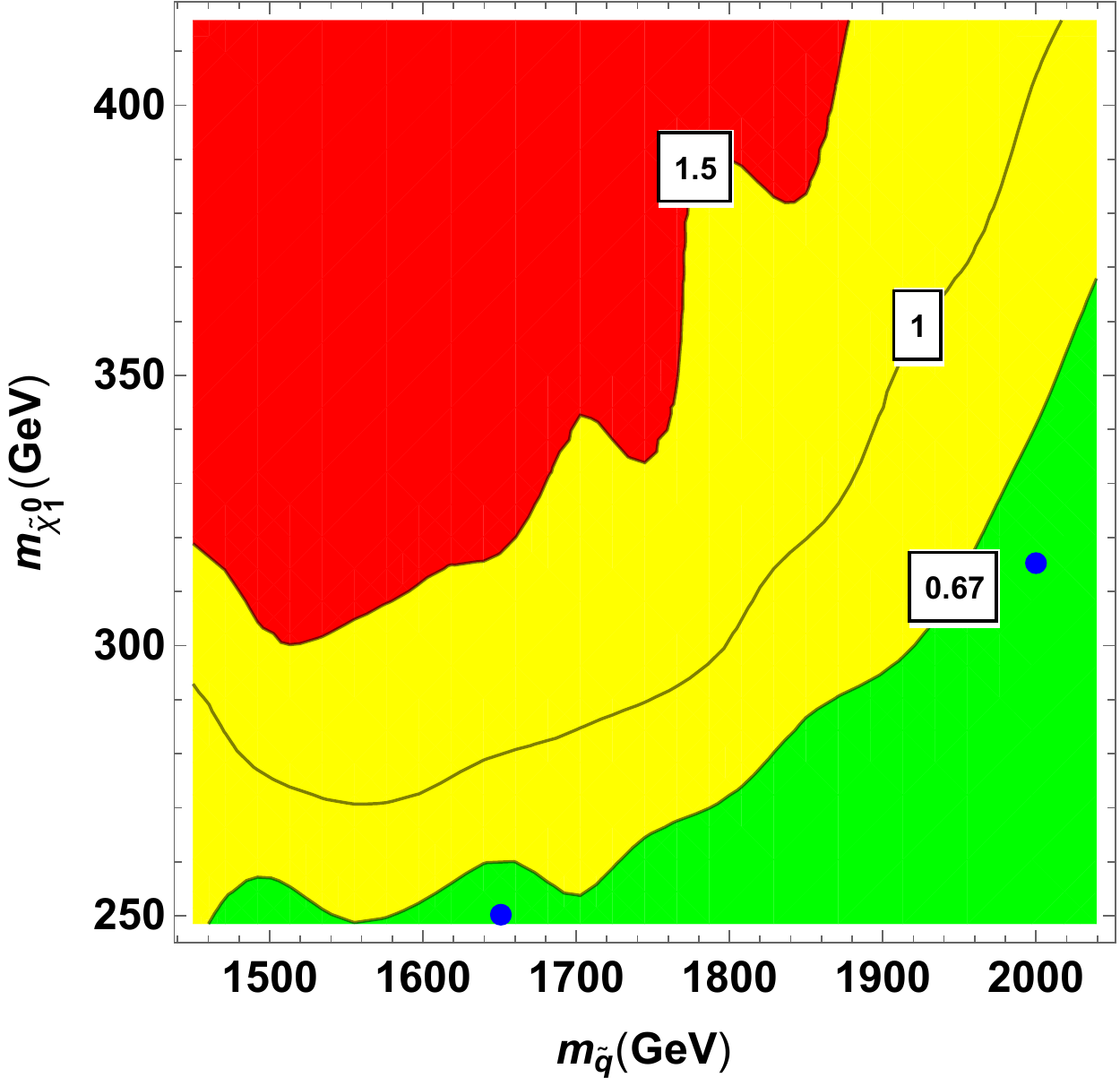}
\caption{Interpolated exclusion plot showing which pairs of bino and squark masses are allowed (green), excluded (red) or ambiguous (yellow) within the MC uncertainty at $\sqrt{s} = 13$ TeV. The blue dots correspond to the specific benchmark points picked for the analysis.}
\label{fig:exclusion}
\end{figure}

We show an exclusion plot that determines which pairs of $m_{\tilde{q}}$ and $m_{\tilde{\chi}_1^0}$ are allowed, excluded or ambiguous (shown as the green, red and yellow areas respectively), which can be seen in Fig. \ref{fig:exclusion}.

With the chosen benchmark points in Fig.~\ref{fig:exclusion}, we perform our analyses using the \textsc{Rivet 2.6.0} analysis toolkit \cite{rivet}. The jets are clustered using \textsc{FastJet 3.3.1}'s anti-$k_T$ algorithm \cite{fastjet} having $R = 1.0$. These large-radius jets are then trimmed \cite{trim} with $p_T > 450$ GeV and $|\eta| < 1.5$ (these will be our large-radius jets).

\begin{figure}[t]
\centering\includegraphics[width=1.0\linewidth]{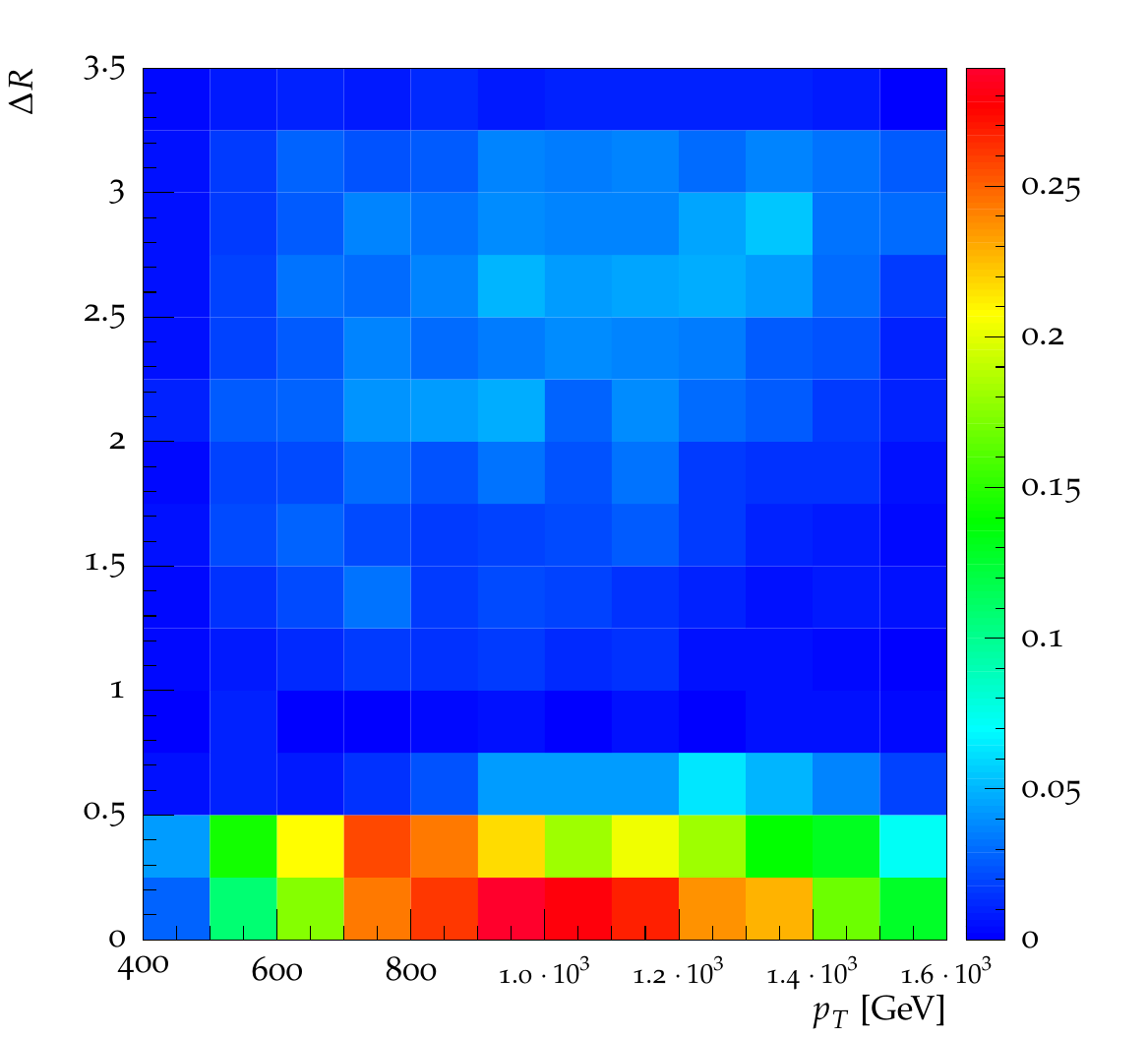}
\caption{This plot shows how the $p_T$ of the leading-mass large-radius jet (which contains the bino decay products) behaves when the $\Delta R$ separation between it and the nearest photon is less than 1. Clearly this shows that the photon is inside the large-radius jet whenever the jet is boosted. This particular plot is for the benchmark point $m_{\tilde{q}} = 1650$ GeV / $m_{\tilde{\chi}_1^0} = 250$ GeV.}
\label{fig:deltaRvsPT}
\end{figure}

Before moving on to study our signal, we made sure that the photon will indeed be contained within the large-radius jet. This of course happens when the topology is boosted. We show this by plotting the $\Delta R$ separation of the nearest photon to the leading-mass large-radius jet versus that jet's momentum. As can be seen in Fig.~\ref{fig:deltaRvsPT}, the photon is inside the large-radius jet (i.e., $\Delta R < 1$) whenever the jet (which contains the decay products of the bino) is boosted.

We then proceeded to study the various kinematic properties of our signals by looking at distributions such as the large-radius jet multiplicities, number of small-radius jets ($R = 0.4$) inside the large-radius jets, $\slashed{E}$ distribution, invariant mass distribution of the leading-$p_T$ and leading-mass large-radius jet, $p_T$ distribution of the leading photon, as well as the $\phi$ distribution between the two leading photons and large-radius jets.

\begin{figure}[t]
\centering\includegraphics[width=1.0\linewidth]{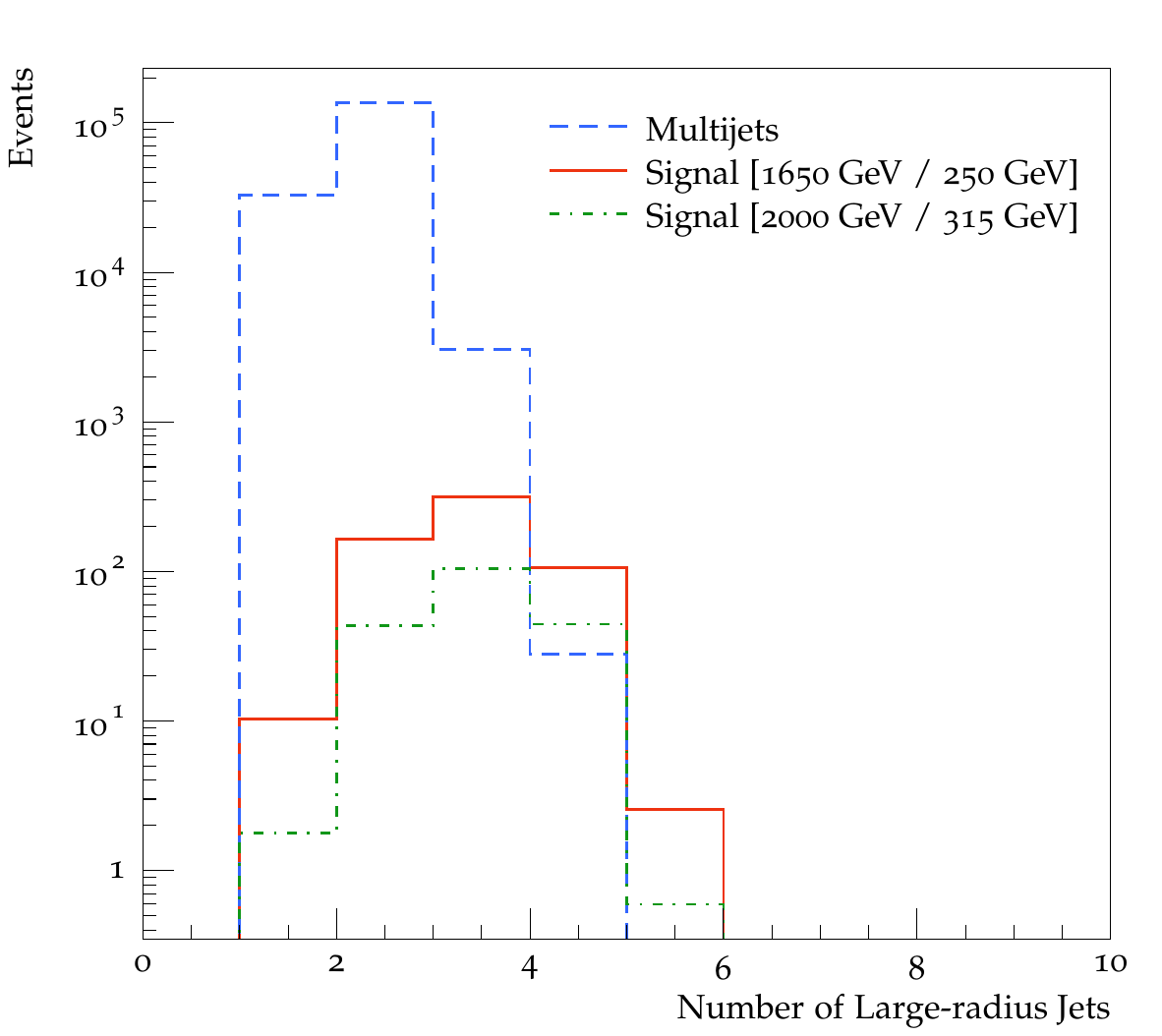}
\caption{Large-radius jet multiplicity after imposing the first two of the selection criteria summarised in Table~\ref{tab:criteria}, for the multijets background and the benchmark point with $m_{\tilde{q}} = 1650$ GeV / $m_{\tilde{\chi}_1^0} = 250$ GeV as well as $m_{\tilde{q}} = 2000$ GeV / $m_{\tilde{\chi}_1^0} = 315$ GeV. Notice how the large-radius jet multiplicity of the signal overcomes that of the background for values greater than 3.}
\label{fig:largejet}
\end{figure}

In the end, we included a lepton veto and then selected events whenever the following requirements are satisfied: (i) the photon (with cuts $p_T > 200$ GeV and $|\eta| < 2.0$) multiplicity is greater than $1$; (ii) the leading-mass large-radius jet ($j_1$) contains its nearest photon ($\gamma$) (i.e., when $\Delta R_{j_1,\gamma} < 1$). We then plot the mass distribution of the leading-mass large-radius jet whenever its large-radius jet multiplicity is greater than $3$. The combination of these three criteria (summarised in Table~\ref{tab:criteria}) turns out to be a strong discriminator against the background (the simulation of which is discussed below) as evidenced by Fig.~\ref{fig:largejet}.

\begin{table}[h]
\centering
\caption{Summary of kinematic cuts and selection criteria}
\begin{tabular}{|l|}
\hline
\begin{tabular}[c]{@{}l@{}}\underline{\textbf{Kinematic cuts}}\\ $\bullet$ Photon: $p_T > 200$ GeV, $|\eta| < 2.0$\\ $\bullet$ Jets: $p_T > 450$ GeV, $|\eta| < 1.5$\end{tabular}                                                                          \\ \hline
\begin{tabular}[c]{@{}l@{}}\underline{\textbf{Selection criteria}}\\ $\bullet$ Photon multiplicity \textgreater 1\\ $\bullet$ $\Delta R_{j_1,\gamma} < 1$\\ $\bullet$ Large-radius jet multiplicity \textgreater 3\end{tabular} \\ \hline
\end{tabular}
\label{tab:criteria}
\end{table}

We also tried using jet substructure variables \cite{substructure1,substructure2} on our large-radius jets to improve the signal over background. These include the LHA (Les Houches angularities) \cite{lha}, Nsubjettiness \cite{nsubjettiness}, ECF (Energy Correlation Function) and $C_2$ (double ratio of ECFs) \cite{ecf}. We found that these substructure variables are not very helpful in our case mainly due to the fact that introducing cuts on these variables reduces our few remaining signals even further after our main cuts have been implemented.

\begin{table*}[]
\begin{tabular}{|l|l|l|l|l|l|}
\hline
                                                            & \begin{tabular}[c]{@{}l@{}}Signal\\ ($0.0127$ pb)\end{tabular} & \begin{tabular}[c]{@{}l@{}}Multijet\\ ($8.70 \times 10^3$ pb)\end{tabular} & \begin{tabular}[c]{@{}l@{}}Photons $+$ jets\\ ($8.067$ pb)\end{tabular} & \begin{tabular}[c]{@{}l@{}}W $+$ jets\\ ($1.397 \times 10^4$ pb)\end{tabular} & \begin{tabular}[c]{@{}l@{}}Z $+$ jets\\ ($6.182 \times 10^3$ pb)\end{tabular} \\ \hline
No Cuts                                                     & 1744                                                           & $1.26 \times 10^9$                                                         & $9.76 \times 10^6$                                                      & $2.09 \times 10^9$                                                            & $9.26 \times 10^8$                                                            \\ \hline
Photon multiplicity \textgreater 1           & 795                                                            & $2.17 \times 10^5$                                                         & $7.96 \times 10^4$                                                      & $2.10 \times 10^3$                                                            & $9.27 \times 10^2$                                                            \\ \hline
$\Delta R_{j_1,\gamma} < 1$                                 & 794                                                            & $2.11 \times 10^5$                                                         & $4.96 \times 10^4$                                                      & $2.10 \times 10^3$                                                            & $9.27 \times 10^2$                                                            \\ \hline
Large jet multiplicity \textgreater 3 & 165                                                            & 99                                                                         & $6$                                                                     & 0                                                                             & 0                                                                             \\ \hline
SR ($200-300$ GeV)                             & 114                                                            & 19                                                                         & 0                                                                       & 0                                                                             & 0                                                                             \\ \hline

\end{tabular}
\caption{Cutflow table showing the number of events for the signal (with benchmark point $m_{\tilde{q}} = 1650$ GeV / $m_{\tilde{\chi}_1^0} = 250$ GeV) and various backgrounds after implementing our main cuts as well as the remaining events in the relevant mass signal region (SR) of $200-300$ GeV. The events are normalised using an integrated luminosity of 150 fb$^{-1}$.}
\label{tab:cutflow}
\end{table*}

Standard Model background comprised of multijets were generated ($5\times 10^6$ events) using \textsc{Pythia 8.235} with a minimum invariant $p_T$ of $300$ GeV. The simulated multijets background has a cross section of $8.53$ nb. We checked that the background event simulation is consistent with matched events from \textsc{MadGraph 2.6.5} \cite{mg5} + \textsc{Pythia 8} as well as \textsc{POWHEG V2} \cite{powheg1,powheg2,powheg3,powheg4} + \textsc{Pythia 8}. We also tested $\gamma+$jets, $W + $ jets, and $Z + $ jets whose contribution to the background turned out to be negligible after all the relevant cuts have been implemented. These can be seen in the cutflow table shown in \ref{tab:cutflow}. Here, the multijets background is clearly the most significant background in the signal region we are looking at. We also investigated  $t\bar{t} + Z$, $t\bar{t} + W$, $Z + \gamma$ as well as $W + \gamma$ background channels but their contributions are negligible as well. For the fake rate, we are only concerned about a jet being misreconstructed as a photon from one of the background samples, thereby pushing an event which would not otherwise count as a real background for lack of a photon in our signal region. Now this is impossible to do at generator level, but one can try to get an estimate from already published analysis what fraction of high $p_T$ photons in data can come from fakes. For example, reference \cite{fakerate} say the effect is at most 10\%.

\begin{figure}[t]
\centering\includegraphics[width=1.0\linewidth]{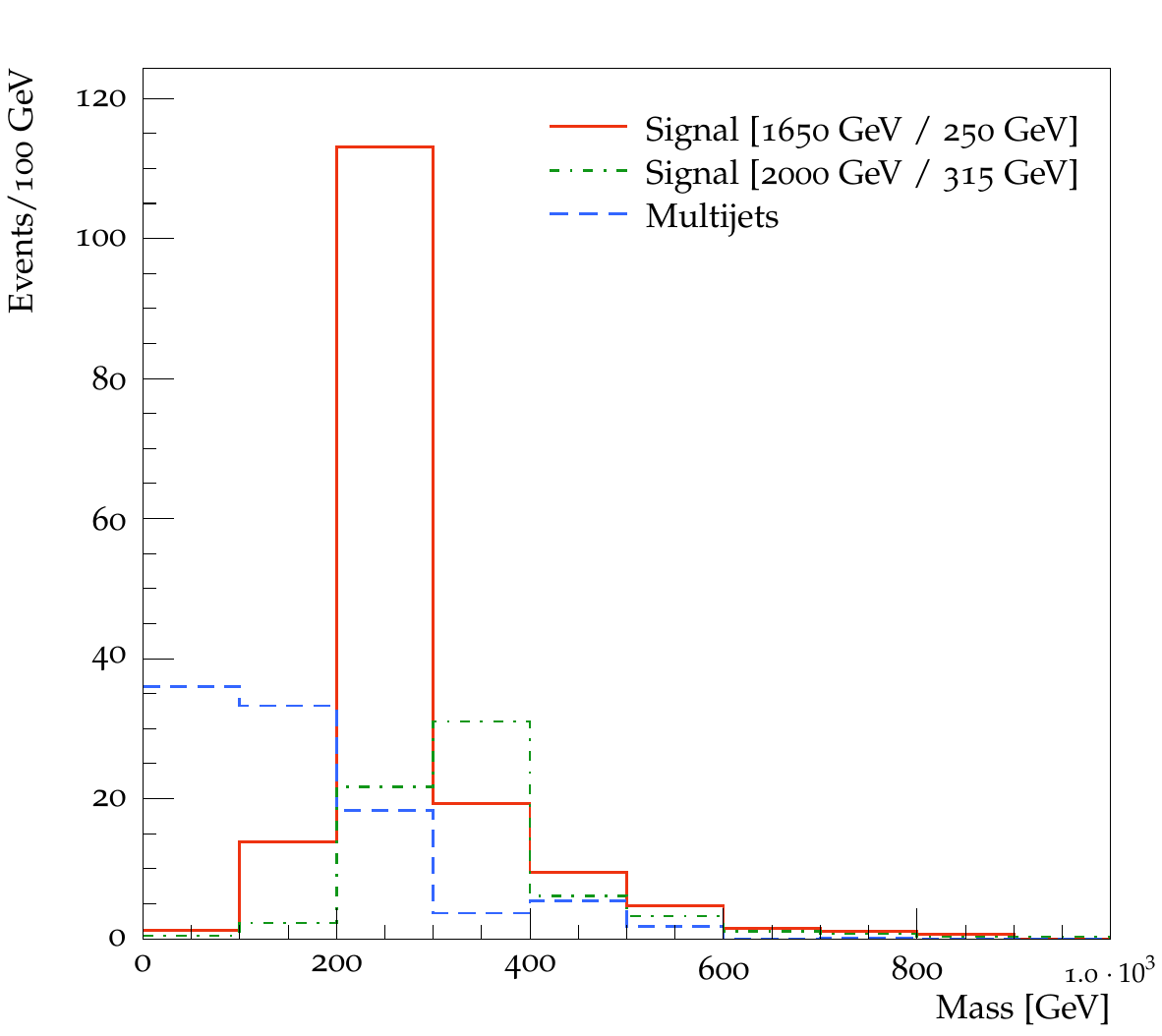}
\caption{Mass distribution of the leading-mass large-radius jet with a photon inside after imposing the selection criteria summarised in Table~\ref{tab:criteria}, using the benchmark point with $m_{\tilde{q}} = 1650$ GeV / $m_{\tilde{\chi}_1^0} = 250$ GeV as well as $m_{\tilde{q}} = 2000$ GeV / $m_{\tilde{\chi}_1^0} = 315$ GeV. Also shown is the multijets background. Note the signal peak at the window where the bino mass is located. All plots are normalised using an integrated luminosity of $150$ fb$^{-1}$.}
\label{fig:1650-250}
\end{figure}

The mass distributions of the benchmark points as well as the background were then normalised to their respective production cross sections using a total integrated luminosity of $150~\mbox{fb}^{-1}$ which roughly corresponds to Run 2 of LHC. Two such distributions are shown in Fig. \ref{fig:1650-250} for the specific benchmark point with $m_{\tilde{\chi}_1^0} = 250$ GeV and $m_{\tilde{q}} = 1650$ GeV as well as $m_{\tilde{\chi}_1^0} = 315$ GeV and $m_{\tilde{q}} = 2000$ GeV. It can be seen that our large-radius jet selection criteria reduces the background below the signal. Even with a realistic reconstruction and trigger efficiency of 50\%, we would still have enough signal events left. An experimentally similar final state, requiring an electron inside a large-radius jet was probed in ATLAS boosted heavy neutrino search \cite{atlaselectron} confirming the feasibility of our method. References \cite{jetphoton1,Domingo:2016unq} also looked at jets formed exclusively from high pT photons.

For the parameter space we have scanned, it turns out that the atlas\_1802\_03158\footnote{This particular search with an integrated luminosity of $36.1$ fb$^{-1}$ is motivated by the gauge-mediated supersymmetric breaking (GMSB) models where the final states that contain large values of $\slashed{E}$ and photons are present. Their search is divided into two regions: (i) diphoton events with large missing transverse energy; (ii) events with missing energy and the presence of one isolated energetic photon. The search is meant to cover gluino, squark and wino/higgsino production and their subsequent decays to NLSP that could decay into a gravitino and a photon or a $Z$ boson.} analysis \cite{experiment1} is always the strongest search. It is clear from Fig.~\ref{fig:1650-250} that our analysis performs better than this particular search for the $m_{\tilde{q}} = 1650$ GeV / $m_{\tilde{\chi}_1^0} = 250$ GeV benchmark point. For the $m_{\tilde{q}} = 2000$ GeV / $m_{\tilde{\chi}_1^0} = 315$ GeV benchmark point however, this is not so clear. Thus we quantify this advantage by comparing the sensitivity of our analysis versus this ATLAS search using $S = s/\sqrt{s+ b}$. For the ATLAS sensitivity, $s$ is the predicted number of signal events while $b$ is the expected Standard Model events quoted by the experiment. For the sensitivity of our analysis, $s$ is still the number of signal events while $b$ is the background events both within the selected large-radius jet mass window of $200-300$ GeV.

\begin{figure}[t]
\centering\includegraphics[width=1.0\linewidth]{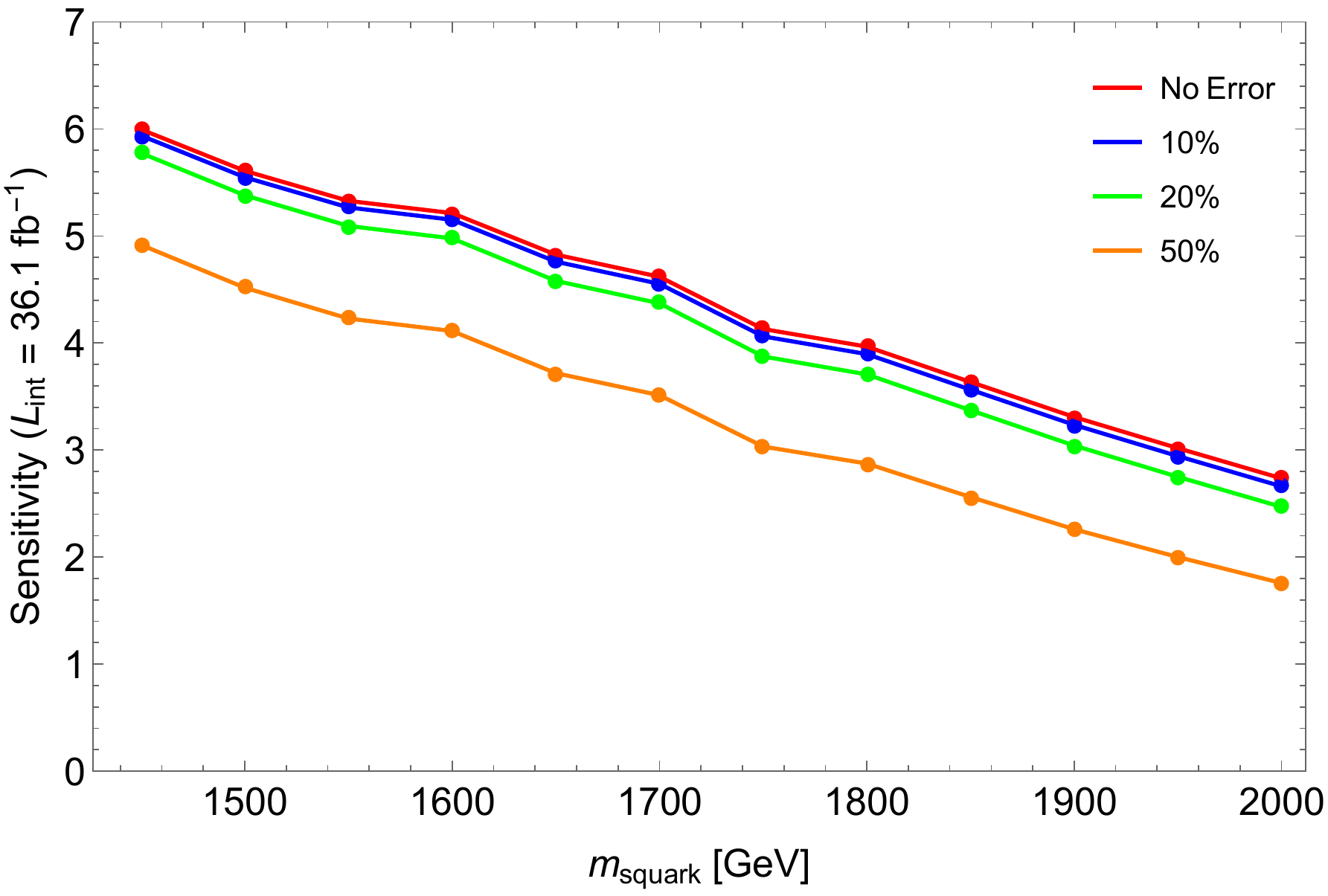}
\caption{Sensitivity plot for various squark masses and a bino mass of $m_{\tilde{\chi}_1^0} = 250$ GeV, scaled down to an integrated luminosity of $36.1$ fb$^{-1}$ for different systematic errors introduced on the background, specifically 0\%, 10\%, 20\%, 50\%.}
\label{fig:sensitivity}
\end{figure}

For the mass distribution shown in Fig.~\ref{fig:1650-250}, the sensitivity (scaled down to correspond to an integrated luminosity of $36.1$ fb$^{-1}$) is given by $4.85$ ($2.58$) for the benchmark $m_{\tilde{q}} = 1650$ GeV / $m_{\tilde{\chi}_1^0} = 250$ GeV ($m_{\tilde{q}} = 2000$ GeV / $m_{\tilde{\chi}_1^0} = 315$ GeV). Even if we scale up our multijet contribution by 10\% from the effect of the fake rate mentioned earlier, our resulting sensitivity is roughly similar. Compare this to the $1.48$ ($0.92$) sensitivity from the ATLAS search also corresponding to an integrated luminosity of $36.1$ fb$^{-1}$. In other words, these benchmark points are clearly excluded by our analysis at $36.1$ fb$^{-1}$. Fig.~\ref{fig:sensitivity} shows our sensitivity for various squark masses, including 10\%, 20\% and 50\% systematic errors introduced on the background. Even with a systematic error of 50\%, the sensitivity of our analysis is still greater than the most sensitive ATLAS search. Since the location of the bino resonance is not determined by SUSY theory, the estimate of significance for our peaks (e.g., in Fig.~\ref{fig:1650-250}) is only the local significance.

\section{Conclusion}
\label{S:4}

To summarise, we have investigated a Stealth SUSY scenario which reduces the missing transverse energy and reconstructed the bino resonances. We looked at various distributions and kinematic properties of our signal such as the large-radius jet multiplicities, number of small-radius jets ($R = 0.4$) inside the large-radius jets, $\slashed{E}$ distribution, invariant mass distribution of the leading-$p_T$ and leading-mass large-radius jet, $pT$ distribution of the leading photon, as well as the $\phi$ distribution between the two leading photons and large-radius jets. In the end, the set of selection criteria that reduces the background below the signal are few and simple. We found that by requiring a high-transverse momentum photon within our large-radius jet, as well as photon and large-radius jet multiplicities to be greater than $1$ and $3$ respectively, we were able to reconstruct the bino mass, $m_{\tilde{\chi}_1^0} \approx M(\gamma gg)$.

We also considered the jet substructure variables of our large-radius jet in order to improve our signal even further but we found that the effects these variables have are not that helpful.

For illustration, we did a simple sensitivity calculation of our signal over background and compared against the strongest existing search and found that our analysis has a better exclusion potential for two benchmark points considered in this letter.

\section*{Acknowledgements}

We thank Xifeng Ruan for an interesting discussion regarding the sensitivity of our search. MF is supported by the University of the Philippines Faculty REPS and Administrative Staff Development Program (FRASDP) and the National Research Foundation - The World Academy of Sciences (NRF-TWAS) Grant No. 110790, Reference No. SFH170609238739. DK is supported by the National Research Foundation of South Africa (Grant Number: 118515).






\begin{thebibliography}{00}


\bibitem{susy} S.P. Martin, Adv. Ser. Dir. High Energy Phys. \textbf{21}, 1 (2010) doi:10.1142/9789814307505\_0001 [arXiv:hep-ph/9709356v7].

\bibitem{dreessusy} M. Drees, R. Godbole and P. Roy, \textit{Theory and Phenomenology of Sparticles}, (World Scientific, 2004).

\bibitem{nillessusy} H. P. Nilles, Phys. Rep. \textbf{110}, 1 (1984) doi:10.1016/0370-1573(84)90008-5.

\bibitem{gunionsusy} J. F. Gunion and H. E. Haber, Nuclear Phys. B \textbf{272}, 1 (1986) doi:10.1016/0550-3213(86)90340-8.

\bibitem{experiment1} The ATLAS Collaboration, Phys. Rev. D \textbf{97}, 092006 (2018), arXiv:1802.03158 [hep-ex].

\bibitem{experiment2} The ATLAS Collaboration, Phys. Rev. D \textbf{97}, 112001 (2018), arXiv:1712.02332 [hep-ex].

\bibitem{experiment3} The ATLAS Collaboration, Eur. Phys. J. C \textbf{76}, 517 (2016), arXiv:1606.09150 [hep-ex].

\bibitem{experiment4} The CMS Collaboration, J. High Energy Phys. \textbf{03}, 166 (2018), arXiv:1709.05406 [hep-ex].

\bibitem{experiment5}  The ATLAS Collaboration, J. High Energy Phys. \textbf{09}, 84 (2017), doi:10.1007/JHEP09(2017)084 [arXiv:1706.03731 [hep-ex]].

\bibitem{Drees:2015aeo}
  M.~Drees and J.~S.~Kim,
  Phys.\ Rev.\ D {\bf 93} (2016) no.9,  095005
  doi:10.1103/PhysRevD.93.095005
  [arXiv:1511.04461 [hep-ph]].
  
\bibitem{Kim:2016rsd}
  J.~S.~Kim, K.~Rolbiecki, R.~Ruiz, J.~Tattersall and T.~Weber,
  Phys.\ Rev.\ D {\bf 94} (2016) no.9,  095013
  doi:10.1103/PhysRevD.94.095013
  [arXiv:1606.06738 [hep-ph]].

\bibitem{Domingo:2018ykx}
  F.~Domingo, J.~S.~Kim, V.~Martin-Lozano, P.~Martin-Ramiro and R.~Ruiz de Austri,
  arXiv:1812.05186 [hep-ph].
  
\bibitem{gambit1} P. Athron \textit{et. al.} (The GAMBIT Collaboration), Eur. Phys. J. C \textbf{77}, 879 (2017), doi:10.1140/epjc/s10052-017-5196-8.

\bibitem{gambit2} P. Athron \textit{et. al.} (The GAMBIT Collaboration), Eur. Phys. J. C \textbf{77}, 824 (2017), doi:10.1140/epjc/s10052-017-5167-0.

\bibitem{fittino1} P. Bechtle \textit{et. al.}, Eur. Phys. J. C \textbf{66}, 215 (2010), doi:10.1140/epjc/s10052-009-1228-3.

\bibitem{fittino2} P. Bechtle \textit{et. al.}, J. High Energy Phys. \textbf{6}, 098 (2012), doi:10.1007/JHEP06(2012)098 [arXiv:1204.4199].

\bibitem{mastercode} K. J. de Vries (MasterCode Collaboration), Nuclear \& Particle Phys. Proc. \textbf{273}, 528 (2016), doi:10.1016/j.nuclphysbps.2015.09.078.

\bibitem{Bertone:2015tza} G.~Bertone, F.~Calore, S.~Caron, R.~Ruiz, J.~S.~Kim, R.~Trotta and C.~Weniger, 
  JCAP {\bf 1604} (2016) no.04, 037
  doi:10.1088/1475-7516/2016/04/037
  [arXiv:1507.07008 [hep-ph]].


\bibitem{experiment6} The ATLAS Collaboration, J. High Energy Phys. \textbf{09}, 88 (2017), doi:10.1007/JHEP09(2017)088 [arXiv:1704.08493 [hep-ex]].


\bibitem{stealth} J. Fan, M. Reece, and J.T. Ruderman, J. High Energy Phys. \textbf{11}, 012 (2011) doi:10.1007/JHEP11(2011)012 [arXiv:1105.5135].

\bibitem{stealth2} J. Fan, M. Reece, and J.T. Ruderman, J. High Energy Phys. \textbf{07}, 196 (2012) doi.org/10.1007/JHEP07(2012)196 [arXiv:1201.4875].

\bibitem{stealth3} J. Fan, R. Krall, D. Pinner, \textit{et. al.} J. High Energy Phys. \textbf{07}, 016 (2016) doi.org/10.1007/JHEP07(2016)016 [arXiv:1512.05781].

\bibitem{atlaselectron} The ATLAS Collaboration, arXiv:1904.12679 [hep-ex].

\bibitem{jetphoton1} The ATLAS Collaboration, Phys. Rev. D \textbf{99}, 012008 (2019) doi:10.1103/PhysRevD.99.012008.

\bibitem{Domingo:2016unq} F.~Domingo, S.~Heinemeyer, J.~S.~Kim and K.~Rolbiecki, 
  Eur.\ Phys.\ J.\ C {\bf 76} (2016) no.5, 249
  doi:10.1140/epjc/s10052-016-4080-2
  [arXiv:1602.07691 [hep-ph]].

\bibitem{rpv} M. Hanussek and J. S. Kim, Phys. Rev. D \textbf{87}, 035002 (2013), arXiv:1211.0725 [hep-ph].

\bibitem{Dreiner:1997uz} H.~K.~Dreiner, 
  Adv.\ Ser.\ Direct.\ High Energy Phys.\ {\bf 21} (2010) 565
  doi:10.1142/9789814307505\_0017
  [hep-ph/9707435].

\bibitem{Dercks:2017lfq} D.~Dercks, H.~Dreiner, M.~E.~Krauss, T.~Opferkuch and A.~Reinert, 
  Eur.\ Phys.\ J.\ C {\bf 77} (2017) no.12, 856
  doi:10.1140/epjc/s10052-017-5414-4
  [arXiv:1706.09418 [hep-ph]].

\bibitem{Barbier:2004ez} R.~Barbier {\it et al.}, 
  Phys.\ Rept.\ {\bf 420} (2005) 1
  doi:10.1016/j.physrep.2005.08.006
  [hep-ph/0406039].


\bibitem{Dreiner:2012gx}
  H.~K.~Dreiner, M.~Kramer and J.~Tattersall,
  EPL {\bf 99} (2012) no.6,  61001
  doi:10.1209/0295-5075/99/61001
  [arXiv:1207.1613 [hep-ph]].

\bibitem{Carena:2008mj}
  M.~Carena, A.~Freitas and C.~E.~M.~Wagner,
  JHEP {\bf 0810} (2008) 109
  doi:10.1088/1126-6708/2008/10/109
  [arXiv:0808.2298 [hep-ph]].

\bibitem{Drees:2012dd}
  M.~Drees, M.~Hanussek and J.~S.~Kim,
  Phys.\ Rev.\ D {\bf 86} (2012) 035024
  doi:10.1103/PhysRevD.86.035024
  [arXiv:1201.5714 [hep-ph]].


\bibitem{Bornhauser:2010mw}
  S.~Bornhauser, M.~Drees, S.~Grab and J.~S.~Kim,
  Phys.\ Rev.\ D {\bf 83} (2011) 035008
  doi:10.1103/PhysRevD.83.035008
  [arXiv:1011.5508 [hep-ph]].

\bibitem{cms3} The CMS Collaboration, CMS PAS B2G-18-007.

\bibitem{cms} The CMS Collaboration, Phys. Lett. B, \textbf{719}, 42 (2013) doi:10.1016/j.physletb.2012.12.055 []arXiv:1210.2052].

\bibitem{cms2} The CMS Collaboration, Phys. Lett. B \textbf{743}, 503 (2015) doi:10.1016/j.physletb.2015.03.017 [arXiv:1411.7255].

\bibitem{nmssm} Ellwanger, U. and Teixeira, A.M. J. High Energ. Phys. \textbf{10}, 113 (2014) doi:10.1007/JHEP10(2014)113 [arXiv:arXiv:1406.7221 [hep-ph]].

\bibitem{jongstealth} J.S. Kim, M.E. Krauss, V.M. Lozano and F. Staub, arXiv:1812.09346v1 [hep-ph].

\bibitem{pythia} T. Sj\"{o}strand \textit{et al}., Comput. Phys. Commun. \textbf{191}, 159 (2015) doi:10.1016/j.cpc.2015.01.024 [arXiv:1410.3012 [hep-ph]].

\bibitem{pdf} R. D. Ball \textit{et. al.} NNPDF Collaboration, 	arXiv:1308.0598 [hep-ph].

\bibitem{nnllfast} W. Beenakker, C. Borschensky, M. Kr\"{a}mer, A. Kulesza and E. Laenen, J. High Energ. Phys. \textbf{1612}, 133 (2016) doi:10.1007/JHEP12(2016)133 [arXiv:1607.07741 [hep-ph]].

\bibitem{checkmate} D. Dercks, N. Desai, J. S. Kim, K. Rolbiecki, J. Tattersall and T. Weber, Comput. Phys. Commun. \textbf{221}, 383 (2017) doi:10.1016/j.cpc.2017.08.021 [arXiv:1611.09856 [hep-ph]].

\bibitem{Kim:2015wza} J.~S.~Kim, D.~Schmeier, J.~Tattersall and K.~Rolbiecki, 
  Comput.\ Phys.\ Commun.\ {\bf 196} (2015) 535
  doi:10.1016/j.cpc.2015.06.002
  [arXiv:1503.01123 [hep-ph]].

\bibitem{Drees:2013wra} M.~Drees, H.~Dreiner, D.~Schmeier, J.~Tattersall and J.~S.~Kim, 
  Comput.\ Phys.\ Commun.\ {\bf 187} (2015) 227
  doi:10.1016/j.cpc.2014.10.018
  [arXiv:1312.2591 [hep-ph]].

  Comput.\ Phys.\ Commun.\ {\bf 221} (2017) 383
  doi:10.1016/j.cpc.2017.08.021
  [arXiv:1611.09856 [hep-ph]].

\bibitem{rivet} A. Buckley et al., Comput. Phys. Commun. 1\textbf{84}, 2803 (2013) doi:10.1016/j.cpc.2013.05.021 [arXiv:1003.0694].

\bibitem{fastjet} M. Cacciari, G.P. Salam, G. Soyez, Eur. Phys. J. C, \textbf{72}, 1896 (2012).

\bibitem{trim} D. Krohn, J. Thaler and LT. Wang, J. High Energ. Phys. \textbf{2}, 84 (2010) doi:10.1007/JHEP02(2010)084 [	arXiv:0912.1342 [hep-ph]].

\bibitem{substructure1} A. J. Larkoski, I. Moult and B. Nachman, arXiv:1709.04464 [hep-ph].

\bibitem{substructure2} L. Asquith, \textit{et. al.}, arXiv:1803.06991 [hep-ex].

\bibitem{lha} S. Badger, \textit{et. al.}, arXiv:1605.04692 [hep-ph].

\bibitem{nsubjettiness} J. Thaler and K. V. Tilburg, J. High Energ. Phys. \textbf{3}, 15 (2011) doi:10.1007/JHEP03(2011)015 [arXiv:1011.2268 [hep-ph]].

\bibitem{ecf} A. J. Larkoski, G. P. Salam, and J. Thaler, J. High Energy. Phys. \textbf{6}, 108 (2013) doi:10.1007/JHEP06(2013)108 [arXiv:1305.0007 [hep-ph]].

\bibitem{mg5} Alwall, J., Frederix, R., Frixione, S. \textit{et. al.}, J. High Energ. Phys. \textbf{7}, 79 (2014) doi:10.1007/JHEP07(2014)079 [arXiv:1405.0301].

\bibitem{powheg1} Alioli, S., Hamilton, K., Nason, P. \textit{et. al.} J. High Energ. Phys. \textbf{4}, 81 (2011) doi:10.1007/JHEP04(2011)081 [arXiv:1012.3380 [hep-ph]].

\bibitem{powheg2} P. Nason, J. High Energ. Phys. \textbf{11}, 040 (2004) doi:10.1088/1126-6708/2004/11/040 [arXiv:hep-ph/0409146].

\bibitem{powheg3} S. Frixione, P. Nason and C. Oleari, J. High Energ. Phys. \textbf{11}, 070 (2007) doi:	10.1088/1126-6708/2007/11/070 [arXiv:0709.2092 [hep-ph]].

\bibitem{powheg4} S. Alioli, P. Nason, C. Oleari and E. Re,  J. High Energ. Phys. \textbf{6}, 043 (2010) doi:10.1007/JHEP06(2010)043 arXiv:1002.2581 [arXiv:1002.2581 [hep-ph]].

\bibitem{fakerate} The ATLAS Collaboration, Eur. Phys. J. C \textbf{78}, 102 (2018) doi:10.1140/epjc/s10052-018-5553-2 [arXiV:arXiv:1709.10440 [hep-ex]].

\end{thebibliography}




\end{document}